\documentclass[12pt,preprint,longabstract]{aastex}
\usepackage{epsfig}
\usepackage{natbib}
\usepackage{graphicx}
\usepackage{tabularx}
\usepackage{rotating}
\usepackage{multirow}
\usepackage{lscape}
\usepackage{mathrsfs,amssymb}
\usepackage{amsmath}
\usepackage{color}

\shorttitle{Physical parameter study of W UMa-type contact binaries in NGC 188}
\shortauthors{Chen et al.}

\begin{document}

\title{Physical parameter study of eight W Ursae Majoris-type contact
binaries in NGC 188}

\author{Xiaodian Chen\altaffilmark{1,2},
        Licai Deng\altaffilmark{2},
        Richard de Grijs\altaffilmark{1,3},
        Xiaobin Zhang\altaffilmark{2},
        Yu Xin\altaffilmark{2},
        Kun Wang\altaffilmark{2,4},
        Changqing Luo\altaffilmark{2},
        Zhengzhou Yan\altaffilmark{2,4},
        Jianfeng Tian\altaffilmark{2},
        Jinjiang Sun\altaffilmark{5},
        Qili Liu\altaffilmark{5},
        Qiang Zhou\altaffilmark{5}, and
        Zhiquan Luo\altaffilmark{4}
        }

\altaffiltext{1}{Kavli Institute for Astronomy \& Astrophysics and
  Department of Astronomy, Peking University Yi He Yuan Lu 5, Hai Dian
  District, Beijing 100871, China; chenxiaodian1989@163.com,
    grijs@pku.edu.cn}
\altaffiltext{2}{Key Laboratory for Optical Astronomy, National
  Astronomical Observatories, Chinese Academy of Sciences, 20A Datun
  Road, Chaoyang District, Beijing 100012, China}
\altaffiltext{3}{International Space Science Institute--Beijing, 1
  Nanertiao, Zhongguancun, Hai Dian District, Beijing 100190, China}
\altaffiltext{4}{School of Physics \& Electronic Information, China
  West Normal University, Nanchong 637002, China}
\altaffiltext{5}{Purple Mountain Observatory, Chinese Academy of
  Sciences, Nanjing 210008, China}

\begin{abstract}
We used the newly commissioned 50 cm Binocular Network (50BiN)
telescope at Qinghai Station of Purple Mountain Observatory (Chinese
Academy of Sciences) to observe the old open cluster NGC 188 in $V$
and $R$ as part of a search for variable objects. Our time-series data
span a total of 36 days. Radial-velocity and proper-motion selection
resulted in a sample of 532 genuine cluster members. Isochrone fitting
was applied to the cleaned cluster sequence, yielding a distance
modulus of $(m-M)_V^0=11.35\pm0.10$ mag and a total foreground
reddening of $E(V-R)=0.062\pm0.002$ mag. Light-curve solutions were
obtained for eight W Ursae Majoris eclipsing-binary systems (W UMas)
and their orbital parameters were estimated. Using the latter
parameters, we estimate a distance to the W UMas which is independent
of the host cluster's physical properties. Based on combined fits to
six of the W UMas (EP Cep, EQ Cep, ES Cep, V369 Cep, and---for the
first time---V370 Cep and V782 Cep), we obtain an average distance
modulus of $(m-M)_V^0=11.31 \pm 0.08$ mag, which is comparable with
that resulting from our isochrone fits. These six W UMas exhibit an
obvious period--luminosity relation. We derive more accurate
  physical parameters for the W UMa systems and discuss their initial
  masses and ages. The former show that these W UMa systems have
  likely undergone angular-momentum evolution within a convective
  envelope (W-type evolution). The ages of the W UMa systems agree
  well with the cluster's age.
\end{abstract}

\section{Introduction}

W Ursae Majoris (W UMa) variables are low-mass, so-called
``overcontact'' binary systems, where the Roche lobes of both stellar
components are filled. W UMas share a common convective envelope. Both
components are characterized by rapid rotation, with periods ranging
from $P = 0.2$ d to $P = 1.0$ d. It is straightforward to obtain
complete and high-quality W UMa light curves in a few nights of
observing time on small to moderate-sized telescopes.

Since W UMas are very common in both old open clusters (OCs) and the
Galactic field, they have significant potential as distance
indicators. Approximately 0.1\% of the F-, G-, and K-type dwarfs in
the solar neighborhood are W UMas \citep{Duerbeck84}, while in OCs
their frequency may be as high as $\sim$0.4\%
\citep{Rucinski94}. Although the occurrence frequency of main-sequence
contact binaries in old globular clusters is low, the frequency of
``blue straggler''-type contact binary systems is two to three times
higher there than that in OCs \citep{Rucinski00}. W UMas can reveal
the evolutionary history of their host cluster, since they are thought
to result from dynamical interactions in the cluster. Alternatively,
these systems may represent a possible final evolutionary phase of
primordial binary systems, once these binaries have lost most of their
angular momentum. Since W UMas are more than 4 mag fainter than
Cepheids or RR Lyrae variables, studies of W UMa distances have only
been undertaken for just a few decades. \citet{Rucinski97} used W UMas
as distance tracers for 400 objects from the Optical Gravitational
Lensing Experiment (OGLE) variable-star catalog.

If a reliable (orbital) period--luminosity (PL) relation can be
established for W UMa systems, they could potentially play a similarly
important role as Cepheids in measuring the distances to old
structures in the Milky Way, including those traced by old OCs and the
Galactic bulge. Although distances based on individual W UMas are not
as accurate as those resulting from Cepheid analysis, their large
numbers could potentially overcome this disadvantage. Clusters
represent good stellar samples to study distances, because their
distances can be estimated in a number of independent ways
\citep[e.g.,][]{Chen15}. In our modern understanding, these
  systems are most likely formed through either nuclear evolution of
  the most massive component in the detached phase (A subtype) or
  angular-momentum evolution of the two component stars within a
  convective envelope (W subtype) \citep{Hilditch1988,
    Yildiz13}. \citet{Yildiz14} provided a method to calculate the
  typical evolution timescales of W UMa systems. Armed with accurate
  distances, significantly improved physical parameters (such as
  masses and luminosities) can be determined. In turn, such W UMa
  systems can then be used to constrain the evolution of W UMa systems
  in general.

NGC 188 is an old OC at a distance of $\sim$2 kpc. It contains a large
number of W UMa variables. Seven W UMas near its center were first
found by \citet{Hoffmeister64} and \citet{Kaluzny87}. \citet{Zhang02,
  Zhang04} undertook a detailed survey covering 1 deg$^2$ around the
cluster center and found 16 W UMas. \citet{Branly96} calculated
light-curve solutions for five central W UMas using the
Wilson--Devinney (W--D) code and discussed the W UMa distance in
relation to the cluster distance. \citet{Liu11} obtained orbital
solutions for EQ Cep, ER Cep, and V371 Cep, while \citet{Zhu14}
published similar results for three additional W UMas, i.e, EP Cep, ES
Cep, and V369 Cep.

In this paper, we study all eight previously identified NGC 188 W UMas
based on high-cadence observations obtained over a continuous period
of more than two months, resulting in accurate, self-consistent, and
homogeneous magnitudes and a well-determined average distance, relying
on up to 3000 data points for a single W UMa system. We also establish
a physical relationship between these eight variables and their host
cluster, NGC 188, based on proper motions, radial velocities, and
features in the color--magnitude diagram (CMD). Although Cepheid
variables are among the most useful objects to establish the distance
ladder, the small number of Cepheids in our Galaxy introduces
relatively large statistical errors, while their disparate distances
introduce comparably large systematic errors in distance modulus,
reddening \citep{An07}, and metallicity \citep{Sandage08}. Compared
with bright O- and B-type Cepheids, faint W UMa dwarfs are much more
plentiful. Our aim is to obtain more accurate cluster distances
  than available to date based on our new W UMa observations and,
  consequently, improve the corresponding PL relation. At the same
  time, an important secondary goal is to obtain significantly
  improved W UMa stellar parameters, thus allowing us to better
  constrain the evolution of our cluster W UMa systems as a
  population.

In Section 2, we discuss our observations and the calibration of both
the NGC 188 data and the W UMa properties used in this study. The
light-curve results, as well as the results from CMD fitting,
proper-motion, and radial-velocity selection, and our distance
analysis, are covered in Section 3. We discuss the properties of our
eight sample W UMas, as well as the feasibility of using W UMas as
distance indicators, e.g., based on their period--luminosity--color
(PLC) or PL relations, in Section 4. In Section 5, we summarize our
main conclusions.

\section{Observations and data reduction}

We observed the OC NGC 188 for a total of 36 nights during two
  separate periods---2014 September 28 to 2014 October 7, and 2016
  January 13 to 2016 March 10---using the 50 cm Binocular Network
telescope \citep[50BiN;][]{Deng13} at the Qinghai Station of Purple
Mountain Observatory (Chinese Academy of Sciences). The time-series
light-curve observations in the Johnson $V$ and $R$ bands were
obtained simultaneously using two Andor 2k$\times$2k CCDs. The
telescope's field of view is $20 \times 20$ arcmin$^2$, which is
adequate for covering the central region of NGC 188. Details of the
observations are included in Table \ref{T1.tab}. Preliminary
processing (bias and dark-frame subtraction, as well as flat-field
corrections) of the CCD frames was performed with the standard {\sc
  ccdproc} tasks in {\sc iraf}. Point-spread-function photometry was
extracted using the {\sc daophot ii} package. Based on a comparison
with the NGC 188 $UBVRI$ photometric catalog of \citet{Sarajedini99},
we calibrated the stellar fluxes and atmospheric absorption for a data
set spanning 7 days (2014 October 1--7). The transfer function is
given by
\begin{equation}\label{equation1}
 \begin{aligned}
  V_0=V+a \times (V-R)+b; \\
  R_0=R+c \times (V-R)+d,
  \end{aligned}
\end{equation}
where $V$ and $R$ are the instrumental magnitudes, while $V_0$ and
$R_0$ are the calibrated magnitudes. The highest-quality $VR$ data
were selected for our color--magnitude analysis, while time-series
data were extracted to find and study variable stars.

%\clearpage
\begin{table}
\caption{\label{T1.tab} Observation log}
\begin{tabular}{ccccccc}
   \hline
  Date        &  Frames & Frames & Date        &  Frames  & Date        &  Frames \\
              &  ($V$)  & ($R$)  &             &  ($V$)   &             &  ($V$)   \\
  \hline
  2014 Sep 28 &  142    &   205  &2016 Jan 25 &  108    &  2016 Feb 09 & 107   \\
  2014 Oct 01 &  219    &   314  &2016 Jan 26 &  111    &  2016 Feb 14 & 100   \\
  2014 Oct 02 &  122    &   196  &2016 Jan 27 &  112    &  2016 Feb 17 & 104   \\
  2014 Oct 04 &  204    &   336  &2016 Jan 28 &  50     &  2016 Feb 19 & 106   \\
  2014 Oct 05 &  209    &   297  &2016 Jan 29 &  46     &  2016 Feb 22 &  96   \\
  2014 Oct 07 &  130    &   212  &2016 Jan 30 &  50     &  2016 Feb 24 & 103   \\
  2016 Jan 13 &  116    &        &2016 Feb 02 &  112    &  2016 Feb 25 & 104   \\
  2016 Jan 14 &  107    &        &2016 Feb 03 &  95     &  2016 Feb 27 &  88   \\
  2016 Jan 15 &   98    &        &2016 Feb 04 &  91     &  2016 Feb 28 & 102   \\
  2016 Jan 20 &  110    &        &2016 Feb 05 &  96     &  2016 Feb 29 & 105   \\
  2016 Jan 22 &  107    &        &2016 Feb 06 &  102    &  2016 Mar 05 & 100   \\
  2016 Jan 23 &   82    &        &2016 Feb 07 &  101    &  2016 Mar 07 &  95   \\
  \hline
\end{tabular}
\end{table}

\section{Results}

\subsection{Membership and Color--Magnitude Diagram}

\begin{figure}
\centering
\includegraphics[width=140mm]{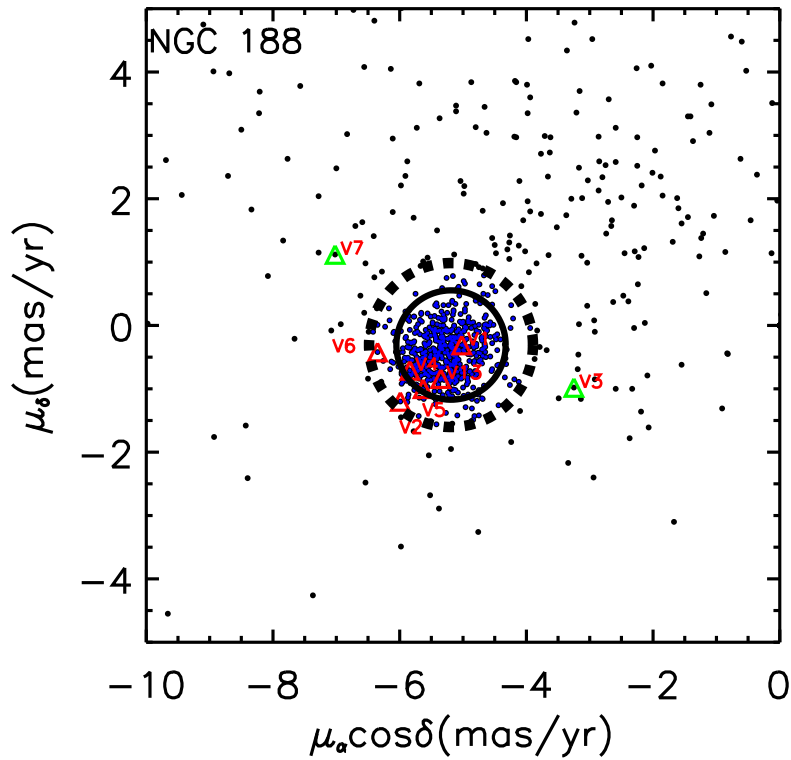}
\caption{Proper-motion distribution of stars in NGC 188
  \citep{Platais03}. The dotted and solid circles represent the
  1.5$\sigma$ and 1$\sigma$ radii, respectively. The blue dots are
  stars located within the (final) $1.5 \sigma$ radius, which are
  treated as genuine cluster members. The triangles represent the W
  UMas' proper motions; red: likely OC W UMas, green: possible OC W
  UMas, located at $\gtrsim 3\sigma$ from the center of the
  distribution.}
  \label{f1.fig}
\end{figure}

Our initial data set contained 914 stars with photometry in both the
$V$ and $R$ filters, and with photometric errors $\sigma_V \le 0.1$
mag for $V \le 18$ mag. We further refined the data set, aiming to
only select genuine cluster members, using proper-motion and
radial-velocity selection. The proper-motion data were obtained from
\citet{Platais03}. Their data base contains proper motions and
positions for 7812 objects down to $V = 21$ mag in a 0.75 deg$^2$ area
around NGC 188. Of our 914 initial sample objects, proper motions
measurements were available for 910. The locus dominated by the
cluster members in the proper-motion distribution is obvious: see
Fig. \ref{f1.fig}. First, we estimated the time average and standard
deviation for all stars in the distribution so as to exclude high
proper-motion stars, adopting a $1.5 \sigma$ selection cut (see
Fig. \ref{f1.fig}). To exclude stars with very large proper motions
and, thus, to only retain the most probable cluster members, this
process was repeated twice (where the distribution's $\sigma$ was
redetermined after the first selection cut), resulting in an average
cluster proper motion of ($\mu_{\alpha}, \mu_{\delta}$) = ($-5.2 \pm
0.6, -0.3 \pm 0.6$) mas yr$^{-1}$. Of the 910 sample stars with
measured proper motions, 558 are located within the (final) $1.5
\sigma$ distribution and they are thus considered cluster members for
the purposes of this paper. Figure \ref{f1.fig} shows the NGC 188
proper-motion distribution; six W UMas (red triangles) satisfy the
selection criteria adopted and are thus most likely NGC 188 W
UMas. Two additional W UMas, i.e., V$_3$ and V$_7$ (green triangles),
are located within the 3$\sigma$ distribution; they are considered
possible OC W UMas.

The average radial-velocity measurements were taken from
\citet{Geller08}. Their sample includes 1046 stars in a 1 deg$^2$
field centered on NGC 188. Radial velocities are known for 433 of the
914 stars in our initial sample. To exclude stars with very large
radial velocities, we applied a three-step\footnote{A three-step
  procedure resulted in the most stable number of members in the final
  sample; a similar result was obtained in two steps based on the
  tabulated proper motions.} selection procedure to all of these 433
stars, adopting (new) $3 \sigma$ selection cuts each time, and
obtained $v_{\rm{RV}}=-42.32 \pm 0.90$ km s$^{-1}$ for the cluster. A
total of 93 stars were excluded from our final sample, leaving 340
genuine cluster members.

\begin{figure}
\centering
\includegraphics[width=140mm]{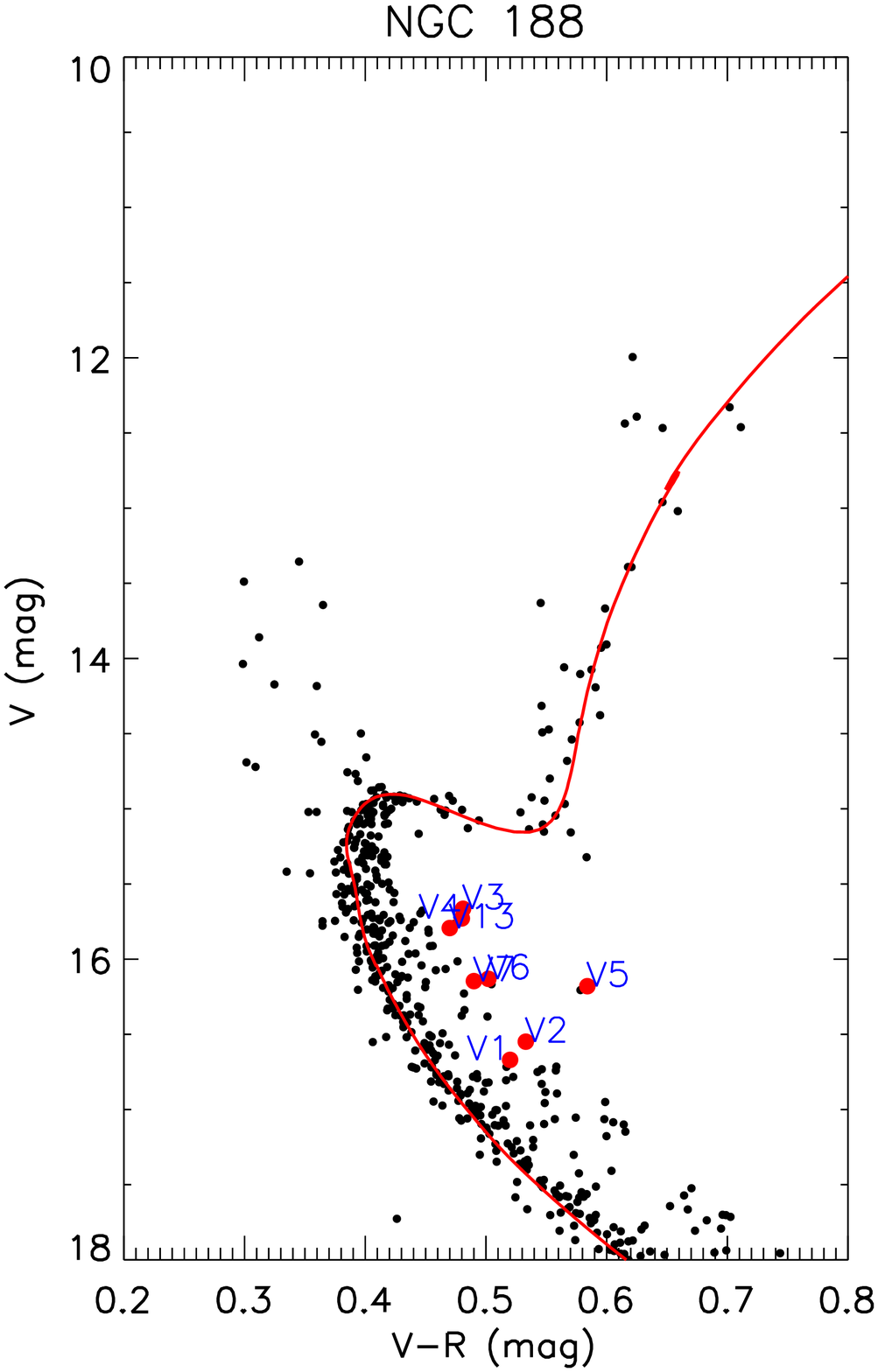}
\caption{Best-fitting isochrone to the CMD of the OC NGC 188. The
  black dots are the genuine cluster members remaining after
  radial-velocity and proper-motion selection. The solid red line is
  the best-fitting isochrone for solar metallicity. The red bullets
  represent the average $(V-R)$ colors and maximum $V$ magnitudes for
  the eight cluster W UMas.}
\label{f2.fig}
\end{figure}

Figure \ref{f2.fig} shows the cluster members' $V,(V-R)$ CMD. After
radial-velocity and proper-motion selection, the cluster members
delineate a clear main sequence, also exhibiting some blue stragglers,
as well as a discernible binary sequence. This old OC has a high
binary frequency of approximately 23\% \citep{Geller12}. We adopted
the Dartmouth stellar isochrones \citep{Dotter08} for further analysis
of the cluster's CMD, since they seem to best match our $VR$
photometry. The Padova isochrones \citep{Girardi00} exhibit a shift in
the locus of the red-giant branch, while the Yale isochrones
\citep{Yi01} cannot be used to match the faint end of the main
sequence ($17 \le V \le 18$ mag). The red line represents the
best-fitting isochrone for an age of 6 Gyr and ${\rm [Fe/H]}=0.0$
dex. We obtained $E(V-R)=0.062 \pm 0.002$ mag and $(m-M)_V^0=11.35 \pm
0.10$ mag. Since the colors of W UMas do not vary significantly during
any given orbital period, we assume $\langle V-R \rangle=\langle V
\rangle-\langle R \rangle$. The loci of the eight W UMas, based on
averaging our photometry over four periods, are shown as red solid
bullets in the CMD, adopting the average maximum magnitudes for each.

\subsection{W UMa light-curve solutions}

\begin{figure*}
\includegraphics[width=160mm]{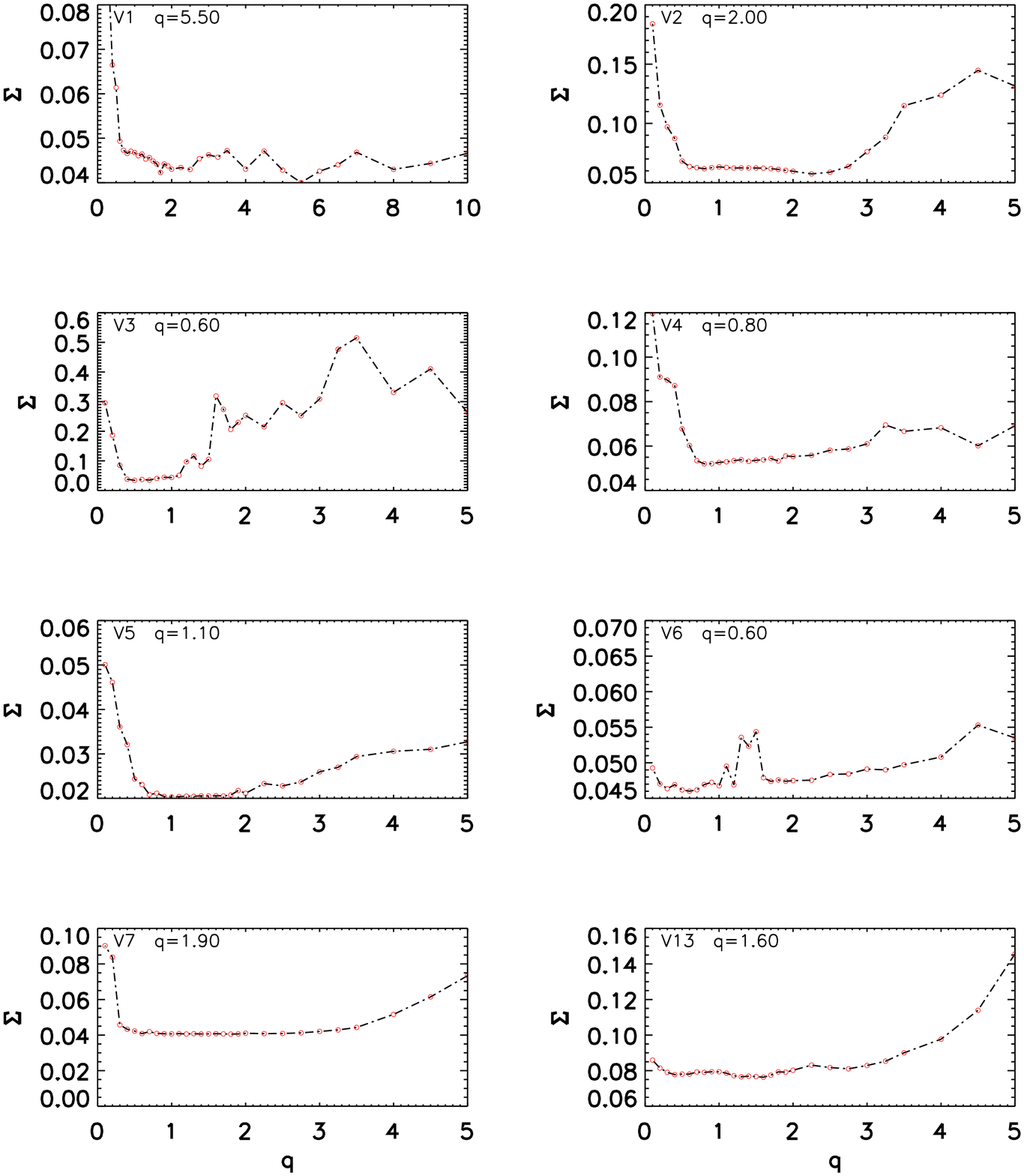}
\caption{Tests to obtain the most appropriate mass ratios, $q$, i.e.,
  those yielding the smallest residuals.\label{f11.fig} }
\end{figure*}

Based on our time-series data, all eight central W UMas were
identified and marked $V_1$--$V_7$ and $V_{13}$ \citep{Zhang04}. We
obtained the best-fitting light-curve solutions for eight W UMa
systems using the same W--D code. First, we can derive the average
temperature $T$, from its intrinsic color. A rough estimate can be
obtained using $T=8540/[(B-V)_0+0.865]$ K. This is usually sufficient,
because its precise value only affects the temperature of the two
components, $T_1$, $T_2$, in the W--D code, while its effect on the
other key parameters, including the mass ratio, stellar radii, and
inclination, is very small. If one is only interested in deriving
orbital parameters, $T$ does not need to be known to very high
accuracy. However, W UMa distance determination depends on knowing a
system's color: see Section 3.4. In this paper, our goal is to
determine temperature-independent distances to our W UMa sample
systems. We therefore need to determine the $(V-R)$ colors more
accurately and apply proper reddening corrections, using
$E(V-R)=0.062$ mag (see Section 3.1). This implies that we need to
obtain more accurate values for $T$, which we can derive from
interpolation of the theoretical color--temperature catalog of
\citet{Lejeune97}. Then, the most important parameter to constrain is
the mass ratio, $q=m_2/m_1$, where $m_1$ and $m_2$ are the masses of
the system's two components, respectively. The mass ratio can be
estimated directly from the radial velocity; $q$ is the inverse of the
ratio of the components' rotational velocities. However, since the
eight W UMas have magnitudes around $V=16$ mag and considering their
rapidly changing velocities, sufficient-quality spectra can only be
obtained using large-aperture telescopes. In addition, a number of
binaries only show a single line in their spectra, so that $q$ cannot
be obtained easily. In the latter case, we can test for
$q\in[0.1,10.0]$ and adopt the mass ratio yielding the smallest
residual as $q$ (see Fig. \ref{f11.fig}). We tested this method
extensively to ascertain that this is a suitable approach to
adequately constrain $q$; this method is most effective for W UMa
systems seen under higher inclinations.

\begin{figure*}
\includegraphics[width=160mm]{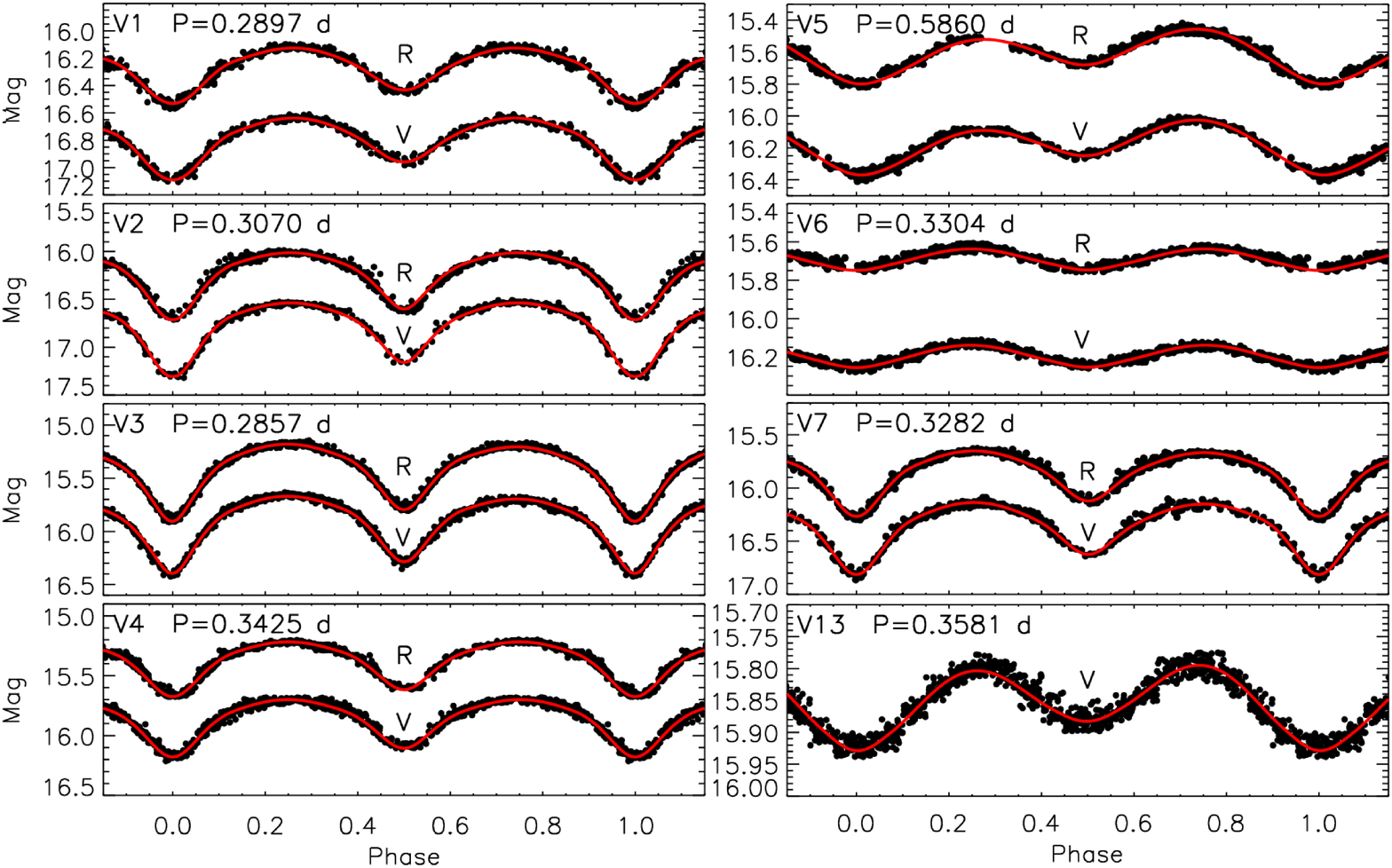}
\vspace{-0.5cm}
\caption{Observations and code solutions of the light curves of eight
  W UMas in NGC 188. The black dots are the observational data, while
  the solid line is the model solution. \label{f4.fig} }
\end{figure*}

The relevant gravity-darkening factor required for application of the
W--D code is usually set at 0.32 and 1.0 for stars dominated by
convective and radiative energy transport, respectively, and the
corresponding reflection factors are 0.5 and 1.0. The bolometric
limb-darkening coefficient for different filters is given by
\citet{van Hamme93}. We also need to adjust $T_2$, the components'
(dimensionless) surface potentials, $\Omega_{1} = \Omega_{2}$, the
primary star's luminosity, $L_1$, and the orbital inclination,
$i$. Figure \ref{f4.fig} shows all observations in phase, with
  the code's best-fitting light-curve solutions overplotted. All eight
  sample W UMas are matched perfectly. Based on assessment of these
light curves, our best fits are significantly better than those
published previously \citep{Branly96,Liu11,Zhu14}, which is largely
driven by the much larger number of observational data points per
single period (and their very homogeneous nature) used for the
fits. Rucinski (1994) used NGC 188 photometry from Kaluzny \& Shara
(1987), which represented no more than 40 data points for a given
period; the scatter associated with their light curves is easily
discernible. The photometric quality used by Branly et al. (1996) was
better than that of Kaluzny \& Shara (1987), at least in the $V$ band
(the $B$-band quality was insufficient for reliable light-curve fits).

%\clearpage
\begin{sidewaystable}
%\hspace{-8cm}
 \footnotesize
  \begin{center}
   \caption{Light-curve solutions for the eight W UMa systems in NGC
     188.\label{T3}}
   \begin{tabular}{lcccccccc}
   \hline
   Parameters               & $V_1$ (EP Cep)& $V_2$ (EQ Cep)& $V_3$ (ER Cep)& $V_4$ (ES Cep)&$V_6$ (V370 Cep)&$V_7$ (V369 Cep)&$V_5$ (V371 Cep)&$V_{13}$ (V782 Cep) \\
   \hline
   $V_{\rm max},R_{\rm max}$ (mag) & 16.670,16.150& 16.549,16.016& 15.665,15.184& 15.729,15.249& 16.131,15.629& 16.146,15.656&16.046,15.470&15.793\\
   $V_{\rm min},R_{\rm min}$ (mag) & 17.084,16.543& 17.301,16.731& 16.409,15.889& 16.140,15.646& 16.245,15.748& 16.808,16.269&16.390,15.803&15.931\\
   $V_{\rm avg},R_{\rm avg}$ (mag) & 16.811,16.286& 16.776,16.235& 15.887,15.398& 15.878,15.393& 16.183,15.684& 16.344,15.844&16.180,15.596&15.858\\
   $T$ (K)                     & 5182       & 5083          & 5458          & 5465          & 5313          & 5396    & 4862 & 5526  \\
   $T_1$ (K)                   & 5600       & 5275          & 5505          & 5582          & 5383          & 5546    & 4780 & 5750  \\
   $T_2$ (K)                & $5074\pm17$   &$4975\pm9$    & $5383\pm10$   & $5308\pm14$   & $5175\pm88$  & $5088\pm12$ & $4935\pm33$& $5370\pm69$ \\
   $\Omega_{1} =\Omega_{2}$ & $9.56\pm0.03$ &$5.35\pm0.01$  &$3.09\pm0.01$  & $3.40\pm0.02$ & $2.93\pm0.05$ &$5.01\pm0.02$ &$3.77\pm0.01$&$4.73\pm0.07$\\
   $i$ ($^\circ$)           &$69.45\pm0.32$ &$81.40\pm0.23$ &$79.69\pm0.15$ &$71.07\pm0.14$ &$46.62\pm2.24$ &$74.71\pm0.18$ &$53.86\pm0.70$&$44.14\pm2.72$\\
   $L_{V_1}/(L_{V_1}+L_{V_2})$          &$0.27\pm0.01$  &$0.41\pm0.01$  &$0.66\pm0.01$  & $0.61\pm0.01$ & $0.68\pm0.02$ &$0.47\pm0.02$ &$0.44\pm0.02$&$0.46\pm0.02$\\
   $q=m_2/m_1$              &$5.41\pm0.02$  &$2.09\pm0.01$  &$0.63\pm0.01$  & $0.78\pm0.01$ & $0.52\pm0.02$ &$1.90\pm0.01$ &$1.06\pm0.01$&$1.60\pm0.03$\\
   $r_1$                    &$0.251\pm0.002$&$0.320\pm0.002$&$0.424\pm0.001$&$0.398\pm0.003$&$0.432\pm0.013$&$0.334\pm0.003$&$0.384\pm0.003$
   &$0.331\pm0.006$\\
   $r_2$                    &$0.537\pm0.002$&$0.449\pm0.002$&$0.343\pm0.001$&$0.354\pm0.003$&$0.318\pm0.012$&$0.446\pm0.003$&$0.395\pm0.003$
   &$0.415\pm0.006$\\
    $(m-M)_V^0$ (mag)        &$11.382\pm0.117$ &$11.280\pm0.111$ &$10.781\pm0.093$ &$11.188\pm0.092$ &$11.294\pm0.122$ &$11.412\pm0.095$  &$11.563\pm0.101$ &$11.314\pm0.115$\\
    $\theta_1$ ($^\circ$)                        &&&&&&&$32.09$&$23.61$\\
   $\psi_1$ ($^\circ$)                         &&&&&&&$315.28$&$341.08$\\
   $r_{\rm spot,1}$ ($^\circ$)                     &&&&&&&$24.97$&$18.66$\\
   $T_{\rm spot,1}/T$                      &&&&&&&$0.50$&$0.73$\\
   $\theta_2$ ($^\circ$)                       &&&79.22&&&&$16.39$&\\
   $\psi_2$ ($^\circ$)                         &&&232.17&&&&$165.10$&\\
   $r_{\rm spot,2}$ ($^\circ$)                     &&&15.80&&&&$25.77$&\\
   $T_{\rm spot,2}/T$                      &&&0.71&&&&$0.68$&\\
   Residual $\sigma$ (mag)  & 0.040         & 0.058          & 0.035        & 0.050         & 0.046         & 0.041 & 0.021    & 0.076    \\
   Membership               & Y             & Y            & N              & Y             & Y             & Y     & Y    & Y   \\
   \hline
 \end{tabular}
 \end{center}
\end{sidewaystable}

\subsection{Primary stellar masses}

In principle, to determine accurate distances to W UMa systems, we
must have access to two absolute measurements. Since we only have
relative photometric data for our eight NGC 188 W UMas, we need to
indirectly obtain absolute measurements using an approximate
method. Fortunately, W UMas are similar to F-, G-, and K-type
main-sequence stars. For these stellar types, semi-observational,
empirical, and theoretical methods of mass determination are readily
available. \citet{Pinheiro14} compared the mass estimates resulting
from four methods: (1) direct stellar-mass determinations from the
objects' spectra; (2) application of the empirical relation between
stellar mass, effective temperature ($T_{\rm eff}$), surface gravity
($\log g$), and metal abundance; (3) application of the empirical
mass--luminosity relation; and (4) a comparison of the objects'
positions in the CMD with Padova isochrones. They found that their
results from application of all four methods were comparable.

\begin{figure}
\includegraphics{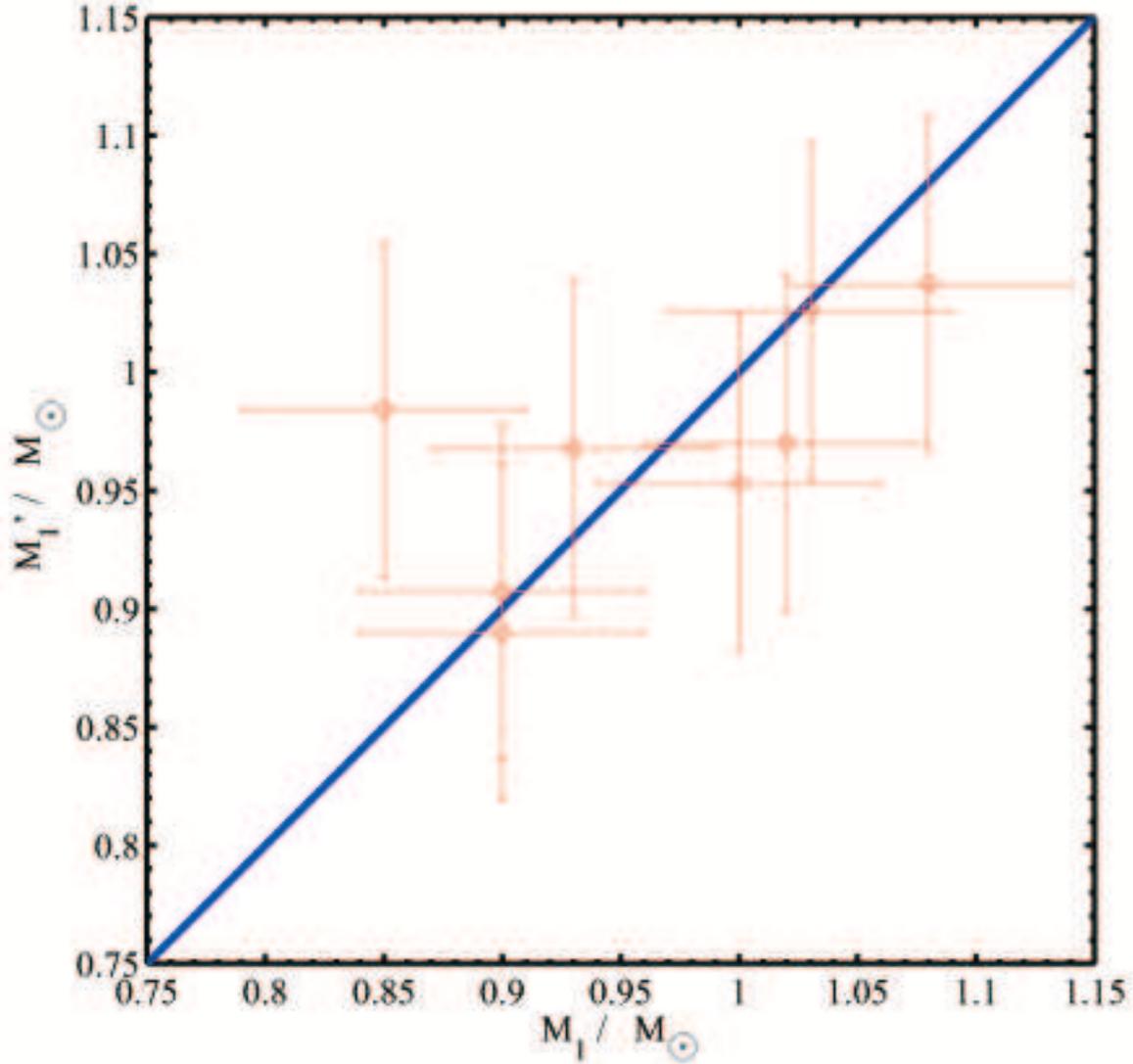}
\caption{Comparison of primary stellar masses based on two different
  methods. The subscripts refer to the methods adopted. $M_1$ is
  estimated from the empirical relationship of \citet{Torres10}, while
  $M_1'$ is estimated from the empirical mass--luminosity relation
  \citep{Henry93}. Considering the prevailing error bars, most mass
  determinations are in agreement, except for V$_5$ (V371
  Cep).  \label{f7.fig}}
\end{figure}

In this paper, we will use methods (2) and (3) to estimate the masses
of the W UMa systems' primary stars. In the context of method (2),
$M_1$ is a function of $f(\log T_1, \log g, {\rm [Fe/H]})$. The
best-fitting results were derived by \citet{Torres10}; the resulting
uncertainty in stellar mass is $\Delta \log(m_1/M_\odot) \simeq 0.027$
for $m_1 = 0.6 M_{\odot}$. $T_1$ is subsequently estimated from the
empirical relationship, and $\log g$ can be obtained from the
best-fitting isochrone. Finally, for NGC 188 we adopt [Fe/H] $= -0.03$
dex \citep{Dias02}. As regards method (3), the primary stellar mass
can be calculated from the absolute magnitude of the W UMas
\citep{Henry93}, while the absolute magnitude can be estimated from
the relevant PL relation (see Section 4.2). The resulting
root-mean-square of the fit is 0.032 in $\log(m_1/M_{\odot})$. In
Fig. \ref{f7.fig}, the difference between the two results is smaller
than the statistical error \citep[cf.][]{Pinheiro14} except for
  V$_5$ (V371 Cep; for a discussion, see Section 4). The primary
stellar mass is not sensitive to the method used, so we use the masses
from \citet{Torres10} to determine the distances to our W UMas.

\begin{figure}
\includegraphics{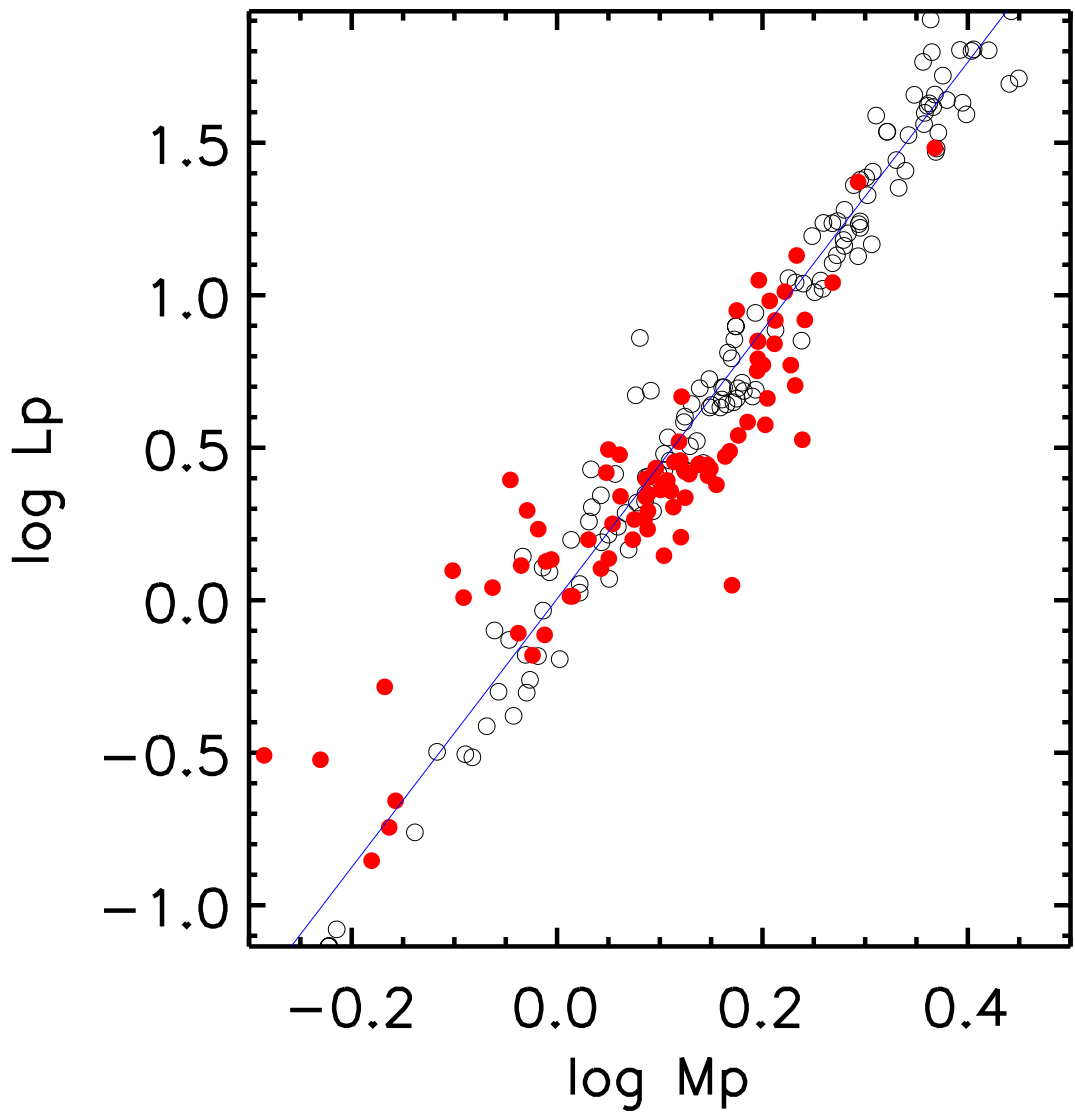}
\caption{Open circles are 190 main-sequence stars in 95 detached
  binaries with mass and radius accuracies within $\pm$3\%
  \citep{Torres10}. The solid bullets are the best-studied 100 W UMas
  of \citet{Yildiz13}. The masses of the solid bullets have been
  corrected onto the main-sequence scale as discussed in the text. The
  solid line represents the $L \propto M^{4.4}$ mass--luminosity
  relation for main-sequence stars. The corrected primary stellar
  masses of the W UMas are well fitted. \label{f8.fig}}
\end{figure}

W UMas are overcontact binary systems. The primary star's mass exceeds
its luminosity, with the excess energy being transferred to its
companion. The correction in mass required from a main-sequence star
to the primary star of a W UMa system is given by $\Delta \log(
m/M_{\odot}) =\frac{1}{4.4}\log (1+U)$, where $U$ is the fraction of
energy of the primary star transferred to the secondary star
\citep{Mochnacki81}. This correction is needed to compare the masses
of main-sequence stars in detached binaries and of primary stars in W
UMas: see Fig. \ref{f8.fig}.

\subsection{Distance determination}

\citet{van Hamme85} provided a method to estimate the absolute stellar
mass from the observed parallax. Once the distance and luminosity are
known sufficiently well, a suitable temperature indicator---such as
the $(B-V)$ color---can permit calculation of a star's absolute
radius, $R$, by application of Stefan's Law. In the Roche model for
binary systems, the semi-major axis, $a$, is related to the size of
the lobe-filling component. The mass ratio can be estimated from the
light-curve solution even if the radial velocity has not been
measured. Both parameters are needed in the application of Kepler's
Third Law to determine the absolute stellar mass, which in turn allows
a distance determination.

\citet{Branly96} derived an equation to estimate the distance based on
$q$ and the relative radius of the primary star, $r_1 = R_1/a$. They
used the maximum $V$ magnitude, $V_{\rm max}$, which is a
superposition of the brightnesses of both component stars. They used
$L_1$ for comparison with $V_{\rm max}$. Following their method, we
use the combination of $L_1+L_2$ instead of $L_1$ alone. From the
fundamental equations
\begin{equation}
\label{equation2}
   \begin{aligned}
   &G(m_1+m_2)=(2\pi/P)^2a^3; \\
   &a=(R/R_{\odot})/r; \\
   &L/L_{\odot}=(T_{\rm eff}/T_{{\rm eff},\odot})^4(R/R_{\odot})^2; \\
   &M_{\rm bol}=M_V+{\rm BC},
   \end{aligned}
\end{equation}
we can derive the distance modulus:
   \begin{equation}
   \label{equation3}
   \begin{aligned}
   V_{\rm max}-M_V=&-39.189+V_{\rm max}+{\rm BC}+10 \log T_1+5 \log r \\
   &           + \frac{5}{3}\log m_1+\frac{10}{3} \log P+\frac{5}{3} \log (1+q);   \\
   &r=\left(r_1^2+r_2^2\left(\frac{T_2}{T_1}\right)^4\right)^{\frac{1}{2}}.\\
   \end{aligned}
   \end{equation}
Here, $G$ is the usual gravitational constant, $M_V$ is the absolute
$V$-band magnitude, $M_{\rm bol}$ is the bolometric luminosity and BC
is the bolometric correction; $r$ is the equivalent relative radius,
which is a combination of $r_1$ and $r_2$, and $m_1$ is expressed in
units of $M_\odot$. The feasibility of using this method depends
almost entirely on knowing the properties of the binary systems
themselves; only the color excess is obtained from fitting the host
OC's isochrones to the observed cluster CMD features. Since the OC's
distance depends only very weakly on its color excess, this method is,
in essence, an independent approach to determine the distances to the
W UMa systems. Table \ref{T3} includes, for all stars, the parameters
that result in the smallest residuals, which in turn can be used to
derive their distances.

To estimate the expected uncertainties pertaining to the resulting
distance modulus, we use
  \begin{equation}
   \label{equation4}
    \begin{aligned}
     \sigma _{V-{{M}_{V}}}^{2}=&\sigma _{V}^{2}+{{\left(
         \frac{\partial \rm BC}{\partial
           (V-R)}+10\frac{\partial \log T_{\rm eff}}{\partial
           (V-R)} \right)}^{2}}\sigma _{V-R}^{2}+\\
    & {{\left( \frac{5}{\ln 10}\frac{1}{r} \right)}^{2}}\sigma
     _{r}^{2}+ \left(\frac{5}{3}\sigma _{{\log
         M}_{1}}\right)^{2}+{{\left( \frac{10}{3\ln 10}\frac{1}{P}
         \right)}^{2}}\sigma _{P}^{2} + \\
    &{{\left( \frac{5}{3\ln 10}\frac{1}{1+q} \right)}^{2}}\sigma
     _{q}^{2} +{{\left( \frac{5}{3\ln 10}
         \right)}^{2}}\frac{6}{r(1+q)}{{\eta }_{rq}}{{\sigma
       }_{r}}{{\sigma }_{q}}, \\
    \end{aligned}
  \end{equation}
where $\sigma_q, \sigma_r$, and $\sigma_P$ are the uncertainties in
the mass ratio, equivalent relative radius, and orbital period,
respectively; $\eta_{rq}$ is the correlation coefficient of $r$ and
$q$. In this equation, $\sigma _{V}$ includes the photometric error
(0.03--0.06 mag, depending on the filter considered), the uncertainty
in our determination of the maximum magnitude (0.002--0.004 mag), as
well as that in our extinction estimate (0.01 mag); $\sigma _{V-R}$
encompasses the photometric error as well as the uncertainty in the
reddening correction (0.002 mag). Since $(V-R)$ is estimated based on
the average value of hundreds of data points, the statistical
uncertainty is negligible, at only 0.0001--0.0002 mag. An uncertainty
of $\sigma _{V-R} = 0.022$--0.042 mag corresponds to an error of
$\sigma_{T_1} = 100$--200 K in our calculation.

The (partial) derivative of BC (and $\log T_{\rm eff}$) to $(V-R)$,
i.e., ${{\left(\frac{\partial \rm BC}{\partial(V-R)}\right)}}$, can be
estimated from interpolation of the equation, and $\sigma_{m_1}$ can
be derived from \citet{Torres10}. If the orbital solution is good
enough, $\sigma _{V}$ and $\sigma _{V-R}$ represent the dominant
components of $\sigma _{V-{{M}_{V}}}$. A prominent error in the
orbital parameters, e.g., in $P, q$, or $T_2$, will increase the
corresponding error. The uncertainties affecting our estimates are all
listed in Table \ref{T4.tab}. The maximum error is approximately 0.3
mag, although uncertainties of 0.1--0.2 mag are expected more
commonly. We thus conclude that W UMa systems can be used to determine
(individual) distances with an accuracy of (often significantly)
better than 0.2 mag in distance modulus, provided that we have access
to high-quality photometry and perform a careful and detailed
analysis.

%\clearpage

\begin{table*}
 \begin{minipage}{180mm}
\caption{Individual uncertainties (in mag) contributing to the total
  error in the distance modulus.}
\label{T4.tab}
\begin{tabular}{lcccccccc}
   \hline
  ID       &$\sigma$   & $\sigma_{V}$  & $\sigma_{(V-R)}$ & $\sigma_{\log m_1}$ & $\sigma_{P}$ & $\sigma_{r}$ & $\sigma_{q}$  & $\sigma_{\rm max}$   \\
  \hline
  EP Cep   &0.12    & 0.04       & 0.08        & 0.04          &   0.006   &0.02       &0.02            & 0.25           \\
  EQ Cep   &0.11    & 0.04       & 0.08        & 0.04          &   0.005   &0.02       &0.02            & 0.23           \\
  ER Cep   &0.09    & 0.03       & 0.06        & 0.04          &   0.004   &0.02       &0.02            & 0.20           \\
  ES Cep   &0.09    & 0.03       & 0.06        & 0.04          &   0.004   &0.02       &0.02            & 0.20           \\
  V370 Cep &0.12    & 0.03       & 0.06        & 0.04          &   0.004   &0.04       &0.08            & 0.26           \\
  V369 Cep &0.10    & 0.03       & 0.06        & 0.04          &   0.004   &0.04       &0.02            & 0.20           \\
  V371 Cep &0.10    & 0.03       & 0.06        & 0.04          &   0.004   &0.05       &0.05            & 0.22           \\
  V782 Cep &0.12    & 0.03       & 0.06        & 0.04          &   0.004   &0.04       &0.07            & 0.25           \\
  \hline
\end{tabular}
\end{minipage}
\end{table*}

\section{Discussion}

\subsection{Individual W UMa systems}

Most previously publications treat $V_3$ (ER Cep) as a cluster
member. This may be owing to the relatively small cluster distance
moduli which were commonly measured prior to 1990
\citep[e.g.,][]{Eggen69,vandenBerg85} or because of a lack of
proper-motion and radial-velocity data. \citet{Worden78} determined an
inclination of $i=79^{\circ} \pm 3^{\circ}$ and a mass ratio $q=0.55
\pm 0.20$. They also estimated a distance modulus of $(m-M)_V^0=10.70
\pm 0.04$ mag, assuming $m_1=1.0 M_{\odot}$, and attempted to use ER
Cep to study the evolution of NGC 188. These authors point out that ER
Cep is located just below the Hertzsprung gap in the CMD, so that the
system could potentially be used to assess the validity of theories
attempting to explain the relevant gap physics, provided that one's
mass accuracy is sufficient. Our results, $q=0.62\pm0.01$ and
$(m-M)_V^0= 10.78 \pm 0.09$ mag, are in agreement with those of
\citet{Worden78}. However, ER Cep's proper motion, ($\mu_{\alpha},
\mu_{\delta}$) = ($-3.25 \pm 0.19, -0.98 \pm 0.19$) mas yr$^{-1}$,
deviates more than 3$\sigma$ from the cluster's mean proper motion,
($\mu_{\alpha}, \mu_{\delta}$) = ($-5.2 \pm 0.6, -0.3 \pm 0.6$) mas
yr$^{-1}$, which means that its cluster membership is
questionable. Its distance modulus, $(m-M)_V^0=10.78\pm0.09$ mag is
clearly significantly different from that of the cluster as a whole,
$(m-M)_V^0=11.279\pm0.010$ mag \citep{Hills15}.

A detailed check of its light curve shows that $V_5$ (V371 Cep) may
not be a genuine W UMa system, since it exhibits clearly unequal
minima and maxima, as well as a strange color and an unusual
period. Its color is obviously redder than those of the other W UMas,
which results in a lower color-based mass than its true mass: see
Fig. \ref{f7.fig}. The apparent 0.2 mag difference between the primary
and secondary minima implies that this object behaves more like an
EB. \citet{Kaluzny87} also identified an EB-type light curve for this
object, with a similar difference between the primary and secondary
minima \citep{Kaluzny90}. \citet{Rucinski97} treated V371 Cep as a
poor thermal-contact or semidetached system and pointed out that this
star is approximately 0.8 mag fainter than expected based on the PLC
relation for W UMas. \citet{Liu11} did not identify the unequal minima
in their light curves, which may be owing to calibration problems
affecting their three nights of observations. To obtain a good fit and
following the commonly adopted approach, we added one spot to both the
primary and secondary stars (see Table \ref{T3}). We thus derived a
distance modulus of $(m-M)_V^0=11.56 \pm 0.10$ mag, which deviates
from the distance obtained from the other W UMas. This thus shows that
V371 Cep is not a well-behaved W UMa-type cluster member.

The orbital-parameter solutions for EQ Cep and ER Cep are comparable
to those determined by \citet{Liu11}. Our ES Cep and EP Cep results
are also comparable with literature determinations, particularly with
those of \citet{Zhu14}, considering that the systematic uncertainty in
the mass ratio is approximately $\Delta q_{\rm syst}=0.1$--0.2 in the
absence of any radial-velocity information. For V369 Cep, we obtain a
larger $q=1.90$, which is in accordance with \citet{Branly96}. In
fact, in their $\Sigma-q$ diagram, $\Sigma$ does not show any
difference for $q\in[0.6, 2.0]$. We find that $q=1.90$ is only 0.5\%
better than other $q$ values in this range. \citet{Branly96} obtained
similar results for their five W UMas (EP Cep, ER Cep, EQ Cep, ES Cep,
and V369 Cep) as we do, despite the differences in our approaches.

The orbital parameters of V370 Cep and V782 Cep have been determined
for the first time here. Since both exhibit small amplitudes of around
0.1 mag, high-accuracy observations are needed to obtain obvious W
UMa-type light curves. We therefore selected all data points with
photometric errors of less than 0.02 mag for our light-curve
fitting. From Fig. \ref{f4.fig} it follows that our light-curve fit
closely matches the theoretical curve.

\subsection{The distance to NGC 188}

NGC 188 is a well-studied Galactic open cluster, and as such its
distance has been measured by many authors. Small cluster distance
moduli of around $(m-M)_V^0=10.80$ mag were commonly measured prior to
1990 \citep[e.g.,][]{Eggen69,vandenBerg85}. \citet{Branly96} derived
two different distance moduli, $(m-M)_V^0=10.80$ mag and
$(m-M)_V^0=11.40$ mag, based on 5 W UMas. During the last decade, a
larger value for the cluster's distance modulus has increasingly been
preferred. \citet{vandenBerg04} derived $(m-M)_V$ values of
11.22--11.54 mag for NGC 188, depending on the chemical composition
assumed. \citet{Meibom09} determined a distance modulus of $(m-M)_V^0
= 11.24\pm 0.09$ mag based on a single EA-type eclipsing binary system
(EA-type systems are detached or semi-detached binary systems whose
secondary minima are almost non-existent). Recently, \citet{Hills15}
published a detailed analysis of the isochrone-fitting method to
determine the distance to NGC 188. They obtained an accurate average
distance of $(m-M)_V^0=11.279\pm0.010$ mag and
$(m-M)_V^0=11.289\pm0.008$ mag based on observations in 15 filters and
for two different isochrone models. Although their statistical
uncertainty is small, the corresponding systematic error would be
0.05--0.10 mag, which is mainly driven by lingering uncertainties in
both the reddening and the theoretical models. In this paper, we
derived a distance modulus of $(m-M)_V^0=11.35 \pm 0.10$ mag from
isochrone fitting, which is indeed comparable to previous
determinations. We also and independently obtained a distance
  modulus of $(m-M)_V^0=11.35 \pm 0.12$ mag based on a combined
  analysis of seven of the eight W UMas in our sample, not including
  the foreground object $V_3$.

Having carefully considered all uncertainties associated with this
approach, we have thus obtained a distance estimate that is indeed
somewhat better than the distances resulting from the empirical,
formulaic approach of \citet{Rucinski94}. Considering all these
results from independent methods, we support a robust distance
determination to NGC 188 of $1800\pm 80$ pc. The distance error
associated with the W UMa method is comparable with that from
isochrone fitting. NGC 188 is a well-studied cluster, so that the
precision of the isochrone-fitting method is assured. Unfortunately,
fewer than half of the OCs in the DAML02 OC catalog \citep{Dias02} are
suitable for use with the isochrone-fitting method because of the lack
of a prominent main sequence, in addition to an absence of
radial-velocity and proper-motion data. The W UMa method further
developed here can play an important role in determining distances to
these clusters.

The W UMa method does not depend on a cluster's age, so it can provide
constraints to estimates of the cluster age. Clusters can provide
useful clues regarding the formation of W UMas. Although W UMa systems
are very common in stellar populations, their formation mechanism is
still controversial. One possible formation scenario is that they form
as a binary system from two close mass overdensities during the time
of molecular-cloud collapse, before they reach the zero-age main
sequence. In this case, the frequency of W UMas in the field and in
OCs must be similar. An alternative scenario suggests that they are
the result of evolved, detached stars. In this latter scenario, the
system's orbital period becomes shorter over time because of angular
momentum loss caused by magnetic braking \citep{Schatzman62}, which
thus implies that the separation between both components
decreases. The frequency of W UMas in clusters of different ages can
provide clues as to their formation. In a region with a radius of 20
arcmin around the NGC 188 cluster center, the frequency of W UMas is
of order 10 per 1000 stars, which is significantly higher than the
average EB-type binary frequency of about 4 per 1000 stars
\citep{Rucinski94} or that of EA-type binaries of about 2 per 1000
stars. Consequently, it seems that W UMas may indeed be capable of
surviving to intermediate ages of several billion years and thus that
they are more likely the products of the evolution of detached
stars. This suggestion should be confirmed by studying a large sample
of OC W UMas.

\subsection{Comparison with the W UMa PLC Relation}

\citet{Rucinski94} pointed out that W UMa-type binaries follow a
strict (empirical) PLC relationship, $M_V =-4.30 \log P + 3.28(B-V)_0
+ 0.04$ mag ($\sigma =0.17$ mag). This relationship is simple and
suitable for application to most W UMa systems. However, it has been
established in part based on 5 W UMa systems in NGC 188. Therefore,
distances derived based on this PLC relation are not independent for
NGC 188. Here we provide a distance comparison using the light-curve
solutions. Table \ref{T5.tab} shows the two distances, while for the
PLC relation we adopt the same color and period as used for the
determination of the light-curve solution distance. Both distances are
in good mutual agreement. For the average distance, the light-curve
solution distance seems slightly better. This may be owing to the
influence of the mass ratio, $q$, which is not included in the
construction of the PLC relation but it is considered in the context
of the light-curve solution distance. \citet{Rucinski97} evaluated the
influence of $q$ and found that it has a $\sigma=0.06$ mag effect on
the PLC relation's zero point. However, more open cluster W UMas with
homogeneous photometry are needed to draw stronger conclusions. In
addition, the good agreement of the two distance measures implies that
we can, in principle, constrain the coefficients of the empirical PLC
relation more tightly.

\begin{table}
\begin{center}
\caption{Comparison of distance moduli based on the two different
  methods employed in this paper.\label{T5.tab}}
\begin{tabular}{lcc}
   \hline
  ID        & $(m-M)_V^0$(This work)        & $(m-M)_V^0$ (PLC)      \\
            & (mag)              & (mag)                                    \\
  \hline
  EP Cep   & $11.382 \pm 0.117$    & 11.502                            \\
  EQ Cep   & $11.280 \pm 0.111$    & 11.383                            \\
  ER Cep   & $10.781 \pm 0.093$    & 10.744                            \\
  ES Cep   & $11.188 \pm 0.092$    & 11.153                            \\
  V370 Cep & $11.299 \pm 0.122$    & 11.341                                   \\
  V369 Cep & $11.412 \pm 0.095$    & 11.425                                   \\
  V371 Cep & $11.563\pm0.101$    & 11.837                                   \\
  V782 Cep & $11.314 \pm 0.115$    & 11.356                                   \\
  Average (except $V_3$ and $V_5$)  & $11.313 \pm 0.079$    & $11.360\pm 0.117$                          \\
  Average (except $V_3$)            & $11.348 \pm 0.119$    & $11.428\pm 0.210$                          \\
  \hline
\end{tabular}
\end{center}
\end{table}

\subsection{Berkeley 39}

Berkeley 39 is a similar OC to NGC 188. It contains 11 W UMa systems
\citep{Kaluzny93,Mazur99}. Consequently, this cluster and its W UMa
sample can be used to double check the applicability of W UMa distance
estimation. \citet{Kiron11} derived the orbital solutions for seven W
UMas in Berkeley 39. Among these seven W UMas, $V_6, V_7, V_8$, and
$V_{10}$ are main-sequence W UMas, while the remaining three stars are
blue straggler-type W UMas. For the four main-sequence W UMas, we
determine a distance modulus $(m-M)_V^0=13.09\pm0.23$ mag, considering
a reddening of $E(B-V)=0.17$ mag and a metallicity ${\rm [Fe/H]}=-0.2$
dex \citep{Bragaglia12}. This distance modulus is comparable to
previous estimates, i.e., $(m-M)_V^0=12.94\pm 0.26$ mag
\citep{Bragaglia12}, $(m-M)_V^0=13.11\pm 0.31$ mag
\citep{Kaluzny89,Mazur99}, and $(m-M)_V^0=13.20\pm0.21$ mag
\citep{Carraro94}, considering differences in the adopted reddening
corrections. Compared with the results for NGC 188, the uncertainty in
the distance is somewhat larger because Berkeley 39 is 3 mag fainter,
which increases the errors in both $V_{\rm max}$ and $T_1$.

\subsection{Period--luminosity relation}

\begin{figure}
\includegraphics{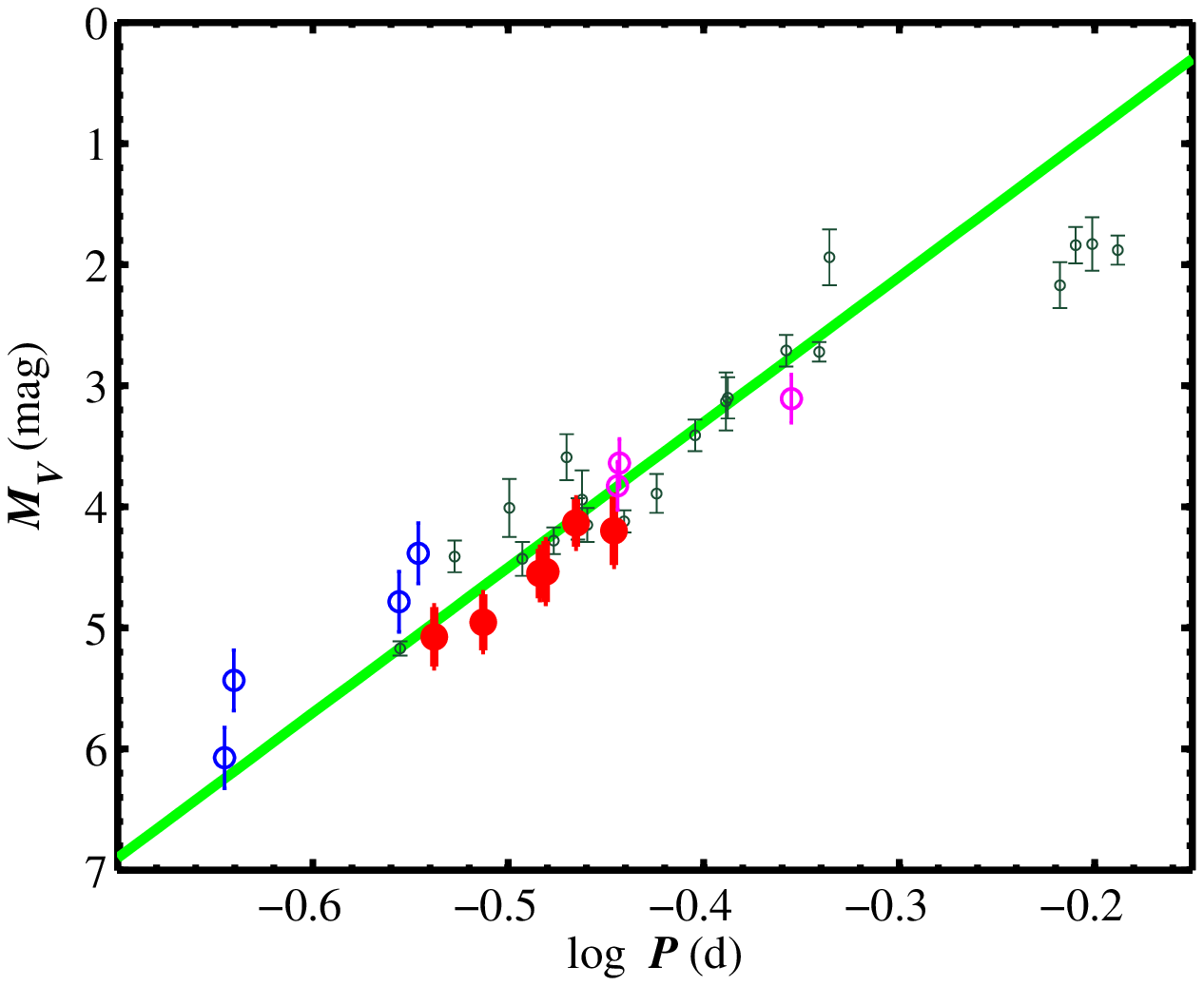}
\caption{$V$-band W UMa PL relation. The heavy red dots are the six
  NGC 188 W UMas retained for our analysis, while the four blue
  circles are main-sequence W UMas in Berkeley 39. The magenta circles
  are three W UMas in M67. The green solid line is the PL relation for
  21 W UMas (small black points with error bars) from
  \citet{Rucinski06}.}
\label{f6.fig}
\end{figure}

\citet{Rucinski06} published a simple $M_V = M_V(\log P)$ calibration,
$M_V =(-1.5 \pm 0.8)-(12.0 \pm 2.0)\log P, \sigma =0.29$ mag, for 21 W
UMa systems with good {\sl Hipparcos} parallaxes and All Sky Automated
Survey (ASAS) $V$-band maximum magnitudes.\footnote{In
  Eq. (\ref{equation3}), the luminosity of the W UMa systems is a
  function of $f(T_{\rm eff},P,q,M,r)$, while their masses and radii
  are known to correlate with $T_{\rm eff}$ \citep{Rucinski94}. In
  principle, the luminosity of W UMas is determined by their intrinsic
  colors, orbital periods, and mass ratios. Since for most W UMas we
  only have access to their orbital periods, it is indeed meaningful
  to establish a simple PL relation to estimate approximate
  distances.} We therefore established the equivalent relationship
using our $V$-band data, combined with the independently determined OC
distance and extinction. We estimated a maximum absolute magnitude for
each W UMa based on distance from \citet{Hills15}, $(m-M)_V^0=11.28$
mag, and reddening, $E(V-R)=0.062$ mag. The results are shown in Table
\ref{T2.tab} and Fig. \ref{f6.fig}. We also added the four W UMas in
Berkeley 39, for comparison, using $(m-M)_0^V = 12.94$ mag,
$E(B-V)=0.17$ mag, and [Fe/H] = $-0.2$ dex \citep{Bragaglia12}.  For
large samples of W UMas, we expect a flat PL relation. This will be
discussed in a subsequent paper, once we have collected sufficiently
large numbers of W UMas spanning a suitably large range in $\log P$.

%\clearpage
\begin{table}
\begin{center}
\caption{Periods and maximum absolute magnitudes for six W
  UMas.\label{T2.tab}}
\begin{tabular}{lcccc}
   \hline
  ID        & $V_{\rm max}$  & $P$      & $M_V$    &  $\sigma_{M_V}$\\
            & (mag)          & (d)      & (mag)    &  (mag) \\
  \hline
  EP Cep   & 16.670    & 0.2897446& 5.144    &   0.25   \\
  EQ Cep   & 16.549    & 0.3069522& 5.023    &   0.23   \\
  ES Cep   & 15.729    & 0.3424570& 4.203    &   0.20   \\
  V370 Cep & 16.131    & 0.3304319& 4.605    &   0.26   \\
  V369 Cep & 16.146    & 0.3281916& 4.620    &   0.20   \\
  V782 Cep & 15.791    & 0.3581278& 4.267    &   0.25   \\
  \hline
\end{tabular}
\end{center}
\end{table}

\subsection{Physical parameters of our sample W UMas}

  The distance modulus to NGC 188 we have determined here matches
  the current best estimate, $(m-M)_V^0=11.28$ mag \citep{Hills15}
  very well. If we now adopt this latter distance modulus, we can in
  turn derive the luminosities, masses, and radii of the W UMa systems
  that are confirmed NGC 188 cluster members. For the foreground W UMa
  system V3, a distance modulus of $(m-M)_V^0=10.70$ mag is
  adopted. The luminosities can be estimated from the systems'
  apparent magnitudes, combined with the adopted distance modulus,
  reliable extinction values, and a proper bolometric correction. The
  radii $R_1,R_2$ can then be derived from their effective
  temperatures, which are known. Since W UMa stars satisfy a
  mass--radius relation, with some corrections $M_1$ follows
  directly. For simplicity, we can straightforwardly adjust the
  semi-major axis, $a$ (i.e., varying only one parameter) in the W--D
  LC code to match the theoretical to the observed luminosity. The
  corresponding $R_1,R_2,M_1$, and $M_2$ for this optimally chosen $a$
  are the resulting masses and radii for the W UMa systems.

The final physical parameters thus derived are included in Table
\ref{T6.tab}. Based on these parameters, V371 Cep is likely a subgiant
or semi-detached contact binary. EP Cep, EQ Cep, V369 Cep, and V782
Cep are obviously W-type W UMa systems, since their less massive
component stars have higher temperatures. ER Cep and V370 Cep are more
likely W-type than A-type W UMa systems, because their temperatures
are lower than those typical of most A-type W UMa systems. It is hard
to classify ES Cep. We have estimated the initial masses based on the
method of \citet{Yildiz13}. Following the criteria proposed by the
latter authors, A-type W UMa systems have $M_{1i}>1.8 M_{\odot}$,
while W-type W UMa systems are characterized by $M_{1i}<1.8
M_{\odot}$. We obtained the same classifications for these W UMa
systems (see Table \ref{T7.tab}). \citet{Yildiz14} proposed a method
to derive the ages of W UMa systems based on their initial masses and
the current masses derived on the basis of the mass--luminosity
relation pertaining to the secondary components. We also calculated
the ages of our seven sample W UMa systems. We found these values to
be in accordance with the overall age of NGC 188 \citep[5--6
  Gyr,][]{Hills15}), except for ER Cep, which is a foreground W UMa
system.

Since we have adopted more accurate distance, better photometric
calibrations, and because we have combined different methods to derive
the masses, our mass determination for each W UMa system is more
accurate than previously published values. \citet{Liu11} obtained
$M_1=0.97 M_{\odot}$ and $M_2=0.51 M_{\odot}$ for EQ Cep, which is
comparable to our mass. However, they used the distance to NGC 188 in
order to calculate the mass of the foreground system ER Cep. As a
consequence, their derived values, $M_1=2.42 M_{\odot}$ and $M_2=1.09
M_{\odot}$, are unrealistic for a W UMa system with a temperature
around 5000 K. \citet{Zhu14} obtained primary masses, $M_1$, of around
0.75 $M_\odot$ for three W UMa systems, i.e., EP Cep, ES Cep, and V369
Cep. These values are underestimates. The corresponding ages of these
W UMa systems, based on these underestimated masses, are close to a
Hubble time, which is not in agreement with the cluster age.

\begin{table*}
\tiny
\caption{Physical parameters of our eight sample W
  UMa systems.\label{T6.tab}}
\begin{tabular}[width=120mm]{lcccccccccc}
   \hline
  ID     & $M_1$& $M_2$ & $R_1$ &$R_2$  & $a$   & $L_1$     & $L_2$ & $T_1$ & $T_2$ & $P$\\
         & ($M_{\sun}$) & ($M_{\sun}$) & ($R_{\sun}$) & ($R_{\sun}$) &($R_{\sun}$) & ($L_{\sun}$) & ($L_{\sun}$) & (K) & (K) & (days) \\
  \hline
   EP Cep & $0.90\pm0.03$& $0.17\pm0.02$ &  $1.01\pm0.04$ & $0.48\pm0.03$  & $1.88\pm0.08$ &  $0.67\pm0.03$  &  $0.22\pm0.10$ &  $5074\pm177$ &  $5600\pm177$ &  0.2897\\
   EQ Cep & $0.90\pm0.03$& $0.43\pm0.02$ &  $0.95\pm0.06$ & $0.68\pm0.05$  & $2.10\pm0.12$ &  $0.62\pm0.05$  &  $0.40\pm0.09$ &  $4975\pm169$ &  $5275\pm169$ &  0.3070\\
   ER Cep & $1.00\pm0.02$& $0.63\pm0.02$ &  $0.91\pm0.03$ & $0.74\pm0.03$  & $2.14\pm0.07$ &  $0.80\pm0.05$  &  $0.48\pm0.08$ &  $5505\pm135$ &  $5383\pm135$ &  0.2857\\
   ES Cep & $1.08\pm0.02$& $0.85\pm0.02$ &  $1.02\pm0.08$ & $0.91\pm0.07$  & $2.56\pm0.20$ &  $1.21\pm0.06$  & $0.79\pm0.13$  &  $5582\pm139$ &  $5308\pm139$ &  0.3425\\
   V371 Cep & $0.85\pm0.03$& $0.80\pm0.03$ &  $1.38\pm0.04$ & $1.35\pm0.03$  & $3.48\pm0.12$ &  $0.95\pm0.10$ &  $0.79\pm0.12$ &  $4935\pm153$ &  $4780\pm153$ &  0.5860\\
   V370 Cep & $1.02\pm0.03$& $0.53\pm0.03$ &  $0.96\pm0.08$ & $0.82\pm0.07$  & $2.32\pm0.18$ &  $0.97\pm0.07$ &  $0.45\pm0.15$ &  $5383\pm213$ &  $5175\pm213$ &  0.3304\\
   V369 Cep & $0.93\pm0.02$& $0.49\pm0.02$ &  $1.01\pm0.08$ & $0.76\pm0.07$  & $2.25\pm0.17$ &  $0.77\pm0.06$ &  $0.61\pm0.08$ &  $5088\pm137$ &  $5546\pm137$ &  0.3282\\
   V782 Cep & $1.03\pm0.03$& $0.64\pm0.03$ &  $1.05\pm0.07$ & $0.84\pm0.06$  & $2.51\pm0.17$ &  $1.06\pm0.11$ &  $0.89\pm0.15$ &  $5370\pm189$ &  $5750\pm189$ &  0.3581\\
  \hline
\end{tabular}
\end{table*}

\begin{table*}
\caption{Initial masses and ages of our seven sample W
  UMa systems.\label{T7.tab}}
\begin{tabular}[width=120mm]{lcccc}
   \hline
  ID     & $M_{2i}$& $M_{1i}$ & $t $ & Type  \\
         & ($M_{\sun}$) & ($M_{\sun}$) & (Gyr) &  \\
  \hline
  EP Cep   &  $1.56 \pm0.06 $& $0.43 \pm0.02$ &  $6.9  \pm1.2 $ & W\\
  EQ Cep   &  $1.42 \pm0.08 $& $0.56 \pm0.03$ &  $7.4  \pm1.6 $ & W\\
  ER Cep   &  $1.07 \pm0.11 $& $0.85 \pm0.09$ &  $12.8 \pm4.2 $ & W\\
  ES Cep   &  $1.76 \pm0.18 $& $0.65 \pm0.07$ &  $5.3  \pm2.8 $ & A(W)\\
  V370 Cep &  $1.29 \pm0.18 $& $0.76 \pm0.11$ &  $8.8  \pm3.7 $ & W\\
  V369 Cep &  $1.52 \pm0.10 $& $0.59 \pm0.03$ &  $5.6  \pm1.3 $ & W\\
  V782 Cep &  $1.46 \pm0.16 $& $0.75 \pm0.08$ &  $5.5  \pm1.9 $ & W\\
  \hline
\end{tabular}
\end{table*}
\section{Conclusions}

Observations of the old OC NGC 188 were obtained with the recently
commissioned 50BiN telescope in $V$ and $R$. We collected 36 nights of
time-series data, spanning an unprecedented total of 3000 frames for
each star. To select genuine cluster members, we performed a detailed
radial-velocity and proper-motion analysis. The radial velocity of NGC
188 is $v_{\rm{RV}}=-42.32 \pm 0.90$ km s$^{-1}$, while its proper
motion is ($\mu_{\alpha},\mu_{\delta}$) = ($-5.2 \pm 0.6, -0.3 \pm
0.6$) mas yr$^{-1}$. Of our total sample of 914 stars, 532 stars are
probable cluster members. They delineate an obvious cluster sequence
down to $V=18$ mag. We use the Dartmouth stellar evolutionary
isochrones \citep{Dotter08} to match the cluster members, adopting an
age of 6 Gyr and solar metallicity. A distance modulus and reddening
of, respectively, $(m-M)_V^0=11.35\pm0.10$ mag and
$E(V-R)=0.062\pm0.002$ mag were obtained.

Accurate light-curve solutions were obtained for the eight W UMas, and
parameters such as their mass ratios and the components' relative
radii were estimated. We subsequently estimated the distance moduli
for the W UMas, independent of the cluster distance. W UMas can be
used to derive distance moduli with an accuracy of often significantly
better than 0.2 mag. Object $V_5$ (V371 Cep) is not a genuine W UMa
system. $V_3$ (ER Cep) was excluded from the distance-modulus analysis
because of its low cluster-membership probability. For the remaining
six OC W UMas---EP Cep, EQ Cep, ES Cep, V369 Cep, V370 Cep, and V782
Cep---we obtained a joint best-fitting distance modulus of
$(m-M)_V^0=11.31 \pm 0.12$ mag, which is comparable to the result from
our isochrone fits, as well as with previous results from the
literature. The resulting accuracy is better than that resulting from
application of the previously established empirical parametric
approximation.

To double check our results for NGC 188 and the applicability of W
UMas as distance tracer, we applied it to the OC Berkeley 39. Based on
four of its W UMas, we derived a distance modulus of
$(m-M)_V^0=13.09\pm0.23$ mag, which is also in accordance with
literature results. W UMas as distance tracers have significant
advantages for poorly studied clusters. The six W Umas in NGC 188
  satisfy a tight PL relation. Armed with the latter, W UMas could
indeed play an important role in measuring distances and to map
Galactic structures on more ambitious scales than done to date.

  Based on better distances and photometry, more accurate physical
  parameters for eight W UMa systems were derived. Using the
  evolutionary W UMa model of \citet{Yildiz13}, the initial masses of
  seven W UMa systems were estimated. All seven W UMa systems have
  initial primary masses below 1.8 $M_\odot$, which means that they
  have evolved along the evolutionar W-type route (i.e., through
  angular-momentum evolution within a convective envelope). The ages
  of seven sample W UMa systems were estimated based on their initial
  masses and the current luminosity-based masses of the secondary
  components. Six of our cluster W UMa systems have similar ages as
  the host cluster itself, which provides confirmation of the
  age-dating method of \citet{Yildiz14}.

%%%%%%%%%%%%%%
\acknowledgments{We are grateful for research support from the
  National Natural Science Foundation of China through grants 11373010
  and 11473037.}

%%%%%%%%%%%%%%%

\end{document}